\documentstyle[epsf,aps]{revtex}
\begin{document}

\def\simge{\,{}^>_{\sim}\,}
\def\simle{\,{}^<_{\sim}\,}

\newcommand{\mbh}{M_{\rm bh}}
\newcommand{\rh}{R_{\rm h}}
\newcommand{\dc}{\delta_{\rm c}}
\newcommand{\mh}{M_{\rm h}}
\newcommand{\ea}{{\em et al.}\,\,}

\draft
\title{Primordial Black Hole Formation 
during First-Order Phase Transitions}
\author{K. Jedamzik}
\address{Max-Planck-Institut f\"ur Astrophysik,
Karl-Schwarzschild-Str. 1, 85740 Garching, Germany} 
\author{J.C. Niemeyer}
\address{University of Chicago, Department of Astronomy and
Astrophysics, 5640 S. Ellis Avenue, Chicago, IL 60637, USA}
\maketitle

\begin{abstract}
Primordial black holes (PBHs) may form in the early universe when
pre-existing adiabatic density fluctuations enter into the
cosmological horizon and 
recollapse. It has been suggested that PBH formation may be facilitated when
fluctuations enter into the horizon during a strongly first-order phase
transition which proceeds in approximate equilibrium.
We employ general-relativistic hydrodynamics numerical simulations in order to
follow the collapse of density fluctuations during first-order phase
transitions. We find that during late stages of the collapse fluctuations separate
into two regimes, an inner part existing exclusively in the high-energy 
density phase with energy density $\epsilon_{\rm h}$,
surrounded by an outer part which exists exclusively in the low-energy 
density phase
with energy density $\epsilon_{\rm h}-L$, where $L$ is the latent heat of the
transition.
We confirm that the fluctuation density threshold 
$\delta\epsilon /\epsilon$ required for the formation
of PBHs during first-order transitions decreases with increasing $L$ and
falls below that for PBH formation during
ordinary radiation dominated epochs. 
Our results imply that, in case PBHs form at all in the early universe,
their mass spectrum is likely dominated by the approximate horizon
masses during epochs when the universe undergoes phase transitions.  
\end{abstract}

\pacs{PACS numbers: 04.70.Bw, 04.25.Dm, 97.60.Lf, 98.80.Cq}

\section{Introduction}

It is long known that only moderate deviations from homogeneity
in the early universe may lead to abundant production of PBHs 
from radiation \cite{PBHorigin}. 
Slightly non-linear fluctuations with excess density
contrast $\delta\epsilon /\epsilon\approx 1$, if horizon-size, are already very
close to their own Schwarzschild radius. Therefore, already moderate collapse
of such fluctuations may lead to the formation of a black hole.
Under such conditions primordial angular momentum of fluctuations
does generally not halt the collapse.
The ultimate fate of an initially super-horizon density fluctuation,
upon horizon crossing,
is mainly determined by a competition between dispersing
pressure forces and the fluctuation's self-gravity.
For an ordinary radiation dominated equation of state (i.e. $p=\epsilon /3$, where
$p$ is pressure) there is approximate
equality between the Jeans-, $M_{\rm J}^{\rm RD}$, 
and horizon-, $M_{\rm h}$, masses.
For fluctuation overdensities exceeding a
critical threshold at horizon crossing $(\delta\epsilon /\epsilon)_{\rm hc} \geq 
\delta_{\rm c}^{\rm RD}\approx 0.7$ 
\cite{Nie1} gravity dominates and the 
formation of a PBH with mass $M_{\rm pbh}\sim M_{\rm h}$ results.
Fluctuations with $(\delta\epsilon /\epsilon)_{\rm hc} 
< \delta_{\rm c}^{\rm RD}$ disperse by
pressure forces.

Since PBH formation from pre-existing adiabatic density fluctuations is
a fine competition between self-gravity and pressure forces any decrease
of the pressure response 
$(\partial p/\partial\epsilon)_{\rm S} = v_{\rm s}^2$ of the radiation
fluid, or equivalently, decrease of the Jeans mass 
$M_{\rm J}\approx (4\pi /3)\epsilon\,\, (v_{\rm s}^2/
4\pi G\epsilon)^{3/2}$, may yield a reduction of the threshold density contrast
$\delta_{\rm c}$ for PBH formation. 
Here $v_{\rm s}$ is the adiabatic speed of sound,
with $v_{\rm s} = 1/\sqrt{3}$ during ordinary radiation dominated epochs.
A decrease in the pressure response of the radiation fluid is, in
fact, anticipated to occur during cosmological first-order phase transitions.
In essence, during a first-order transition between a high-energy density
phase with energy density, $\epsilon_{\rm h}$, and a low-energy
density phase with energy density $\epsilon_{\rm l} = \epsilon_{\rm h} - L$ both
phases may coexist in pressure equilibrium, $p_{\rm h} = p_{\rm l}$ 
at a coexistence
temperature $T_{\rm c}$. In a state where a fluid element is permeated
by both phases, compression leads to an increase of
energy density, since some low-energy density phase 
is converted into high-energy density phase. 
Nevertheless, there is no increase in pressure such that $v_{\rm
s}^{\rm eff}\approx 0$ \cite{Jed1,Schmid1}
for a fluid element substantially larger than the mean separation between
high-energy- and low-energy density phases.
(see \cite{Jed1} for a more detailed discussion).
These considerations are, of course, only valid under the assumption of
approximate maintenance of thermodynamic equilibrium, in particular, 
negligible super-cooling and -heating. Less dramatic reductions of
$v_{\rm s}^{\rm eff}$
may also occur during higher-order phase transitions 
or particle annihilation periods in the early universe.

A reduction of the PBH formation threshold 
for fluctuations which enter the cosmological horizon during first-order
phase transitions (i.e. $\delta_{\rm c}^{\rm FPT} < \delta_{\rm c}^{\rm RD}$)
may have cosmological implications even if it is only modest.
Conversion of cosmic radiation into PBHs at early epochs must be an 
extremely inefficient process if 
the contribution of PBH mass density to the present closure density, 
$\Omega_{\rm pbh}$, is not too exceed unity.
For Gaussian fluctuations this implies
that PBH formation results only from those overdense fluctuations well within
the exponentially declining tail of the density distribution function. 
PBH number density is dominated for fluctuations with 
$(\delta\epsilon /\epsilon )_{\rm hc}$ in a very narrow range between
$\delta_{\rm c}$ and $\delta_{\rm c}+ \sigma^2 /\delta_{\rm c}$, 
where $\sigma$ is the 
variance of the Gaussian distribution. Typically, 
$\sigma^2 /\delta_{\rm c} \simle 10^{-2}$ for $\Omega_{\rm pbh}\simle 1$.
For approximately scale-invariant Harrison-Zel'dovich spectra of the 
primordial density fluctuations, such as resulting from a multitude of
proposed inflationary scenarios, fluctuations re-enter into the horizon
with equal amplitudes on all mass scales. In this case, the slightest
reduction of $\delta_{\rm c}^{\rm FPT}$ as compared to $\delta_{\rm c}^{\rm RD}$
may result in the formation of PBHs on essentially only the horizon scale
during the first-order phase transition, yielding a highly peaked PBH mass
function. We note here that there are other proposed scenarios for PBH formation
during first-order phase transitions \cite{otherPBH}.
These usually involve production of seed fluctuations during the transition
which collapse to PBHs.

The above considerations have led one of us
to propose PBHs formed during the QCD color deconfinement transition
at temperature $T_{\rm c}\approx 100$MeV and
with typical masses $M_{\rm pbh}\sim 
M_{\rm h}^{\rm QCD} \approx 2 M_{\odot} (T/100 {\rm
MeV})^{-2}$ as a candidate for halo dark matter \cite{Jed1,Jed2}.
Compact halo dark matter with approximate masses in the range
$0.1M_{\odot}\simle M\simle 1M_{\odot}$
may have been detected by the MACHO \cite{MACHO} and EROS \cite{EROS}
collaborations. They monitored the light curves of millions of stars within the
Large- and Small- Magellanic Clouds (LMC and SMC), searching for
gravitational microlensing of LMC (SMC) stars by foreground compact objects.
The simplest interpretation of their findings is that a significant fraction of
halo dark matter is within compact objects of unknown nature.
A dynamically significant population of halo red- or white- dwarfs is subject
to stringent constraints from either direct observations
\cite{direct}, or considerations of 
chemical evolution and contribution of MACHO baryon 
density to the critical density \cite{Field}.
Nevertheless, recent observations 
of microlensing of a star in the SMC by a binary lens 
\cite{SMC98-1} may favor a cosmologically
less interesting interpretation of the MACHO/EROS 
observations. It has been
shown that the binary lens responsible for the microlensing
event 98-SMC-1 most likely resides within the SMC itself, and not 
within the galactic halo. 
It may well be
that lenses responsible for other microlensing events are not 
within the galactic halo
but represent so far unknown stellar populations within the LMC
and SMC. Clearly, further observations are needed to reveal
the nature and location of the lens population. Note that a population
of halo PBHs could be revealed by detection of gravitational
waves emitted during PBH-PBH mergers \cite{Naka}.

In this paper we employ a general-relativistic hydrodynamics code
to follow the evolution of density fluctuations which enter the
cosmological horizon 
during a first-order phase transition. In Section 2. we briefly summarize
the adopted numerical technique which has been discussed in detail by
\cite{Baumgarte} and \cite{Nie1} 
and introduce the equation of state describing the phase transition.
Results and discussion of the PBH formation process are presented in Section 3.

\section{The Phase Transition and Numerical Simulation Technique}
\label{numerics}

The algorithm employed to follow the evolution of spherically symmetric
density fluctuations in an expanding Friedman-Robertson-Walker (FRW)
universe is the same as the one adopted and described in detail in
\cite{Baumgarte,Nie1}.
The method, based on an algorithm developed by Baumgarte {\it et al.}
\cite{Baumgarte}
and modified for use in a FRW background, is particularly well suited not
only to simulate the formation of a black hole, but also to
follow the late-time evolution of PBHs by virtue of the chosen Hernandez-Misner
coordinates \cite{hermis66}. 
For details concerning the algorithm, zoning, size of time steps
and computational domain, etc.
the reader is referred to \cite{Nie1}.
Initial conditions for the metric perturbations are chosen to be pure energy
density fluctuations with Mexican-hat profile

\begin{equation}
\label{mexhat}
\epsilon(R) = \epsilon_0 \left[1 + A \left(1 - \frac{R^2}{\rh^2}\right)
\exp{\left( -\frac{3 R^2}{2 \rh^2}\right)}\right] \,\,,
\end{equation}
in unperturbed uniform Hubble expansion (essentially synchronous gauge
with uniform Hubble expansion condition). 
Here, $R$ is circumferential radius
(i.e. $R$ appears in the angular piece of the metric $R^2d\Omega^2$,
where $d\Omega$ is the solid angle element), which in the absence
of perturbations, corresponds to what is
commonly referred to as proper distance in cosmology, $r_{\rm p}=ar_{\rm c}$,
where $a$ is the scale factor of the universe and $r_{\rm c}$ 
is the comoving cosmic
distance. The distance $R_{\rm h}$ is
chosen to be the cosmological horizon distance, $R_{\rm h}(t_0)$, 
at the beginning of the simulation
$t_0$, i.e. 

\begin{equation}
\label{horizon}
R_{\rm h}(t_0) = a(t_0) \int_0^{t_0} {dt\over a(t)} \,\,,
\end{equation}
The increase of horizon distance with cosmic time 
in the presence of a phase transition is computed numerically.
The amplitude $A$ is adjusted
such that fluctuation amplitudes are characterized 
by a certain average overdensity at horizon crossing

\begin{equation}
\label{average}
\biggl({\delta\epsilon\over\epsilon}\biggr)_{\rm hc} \equiv (V_{\rm h}
\epsilon_0)^{-1} 
4\pi\int_0^{R_{\rm h}(t_0)}\bigl(\epsilon (R) -\epsilon_0\bigr)R^2dR
- 1\,\,,
\end{equation}
where $\epsilon_0 =\epsilon_0(t_0)$ is the unperturbed FRW energy density
at $t_0$ and $V_{\rm h} = (4\pi /3) R_h^3(t_0)$ is the initial
horizon volume.

The properties of the equilibrium first-order phase transition are
fully described by a choice for the equation of states of the low- and
high-energy density phases,
respectively. We adopt a phenomenological bag model for the high-energy
density phase, where the energy density is given by the contributions of a
quasi-free, extremely relativistic gas with statistical weight $g_h$ plus a 
temperature-independent self-interaction correction, the bag constant $B$.
We approximate the low-energy density phase as a non-interacting, extremely 
relativistic gas with statistical weight $g_l$. 
Thermodynamic quantities of the individual phases are \cite{Ful}

\begin{equation}
\label{density}
\epsilon_l(T)=g_l{\pi^2\over 30}T^4\quad ;\quad \epsilon_{h}(T)=g_{h}
{\pi^2\over 30}T^4+B\,\ ;
\end{equation}
\begin{equation}
\label{pressure}
p_l(T)={1\over 3}g_l{\pi^2\over 30}T^4\quad ;\quad p_{h}(T)={1\over 3}g_{h}
{\pi^2\over 30}T^4-B\,\ ;
\end{equation}
\begin{equation}
\label{entropy}
s_l(T)={4\over 3}g_l{\pi^2\over 30}T^3\quad ;\quad s_{h}(T)={4\over 3}g_{h}
{\pi^2\over 30}T^3\,\ ,
\end{equation}
where $s$ denotes entropy density. Requiring the existence of a first-order
phase transition at temperature $T_c$, i.e. $p_l(T_c)=p_h(T_c)$, we find
$\epsilon_h(T_c) -  \epsilon_l(T_c) = 4B$, which together with the definition of
latent heat, $L = T_c (\partial /\partial T) (p_h - p_l) = T_c (s_h - s_l)$, implies
$L = \epsilon_h(T_c) -  \epsilon_l(T_c)$. We define the useful quantity

\begin{equation}
\label{eta}
\eta = {L\over \epsilon_l(T_c)} = {{\epsilon_h(T_c) -  \epsilon_l(T_c)}\over
\epsilon_l(T_c)}\,\,,
\end{equation}
describing the strength of the phase transition. Since we assume the
transition to proceed in close-to-equilibrium 
(i.e. negligible super -cooling and -heating) we may use
conservation of entropy to relate the ratio of scale factors 
at the beginning of the transition,
$a_1$, and the end of the transition, $a_2$, to $\eta$

\begin{equation}
\label{length}
{a_2\over a_1} = \biggl({g_h\over g_l}\biggr)^{1/3} = 
\biggl(1+{3\over 4}\eta\biggr)^{1/3}\,\, . 
\end{equation} 
During ordinary radiation dominated epochs the dynamics of fluctuations
is self-similar for fluctuations on small length scales collapsing at early times
and fluctuations on large length scales collapsing at late times, only dependent
on the shape and density contrast of the fluctuation. 
Similarly, fluctuation dynamics in the presence of a phase transition should
be independent of the temperature, energy density, and horizon mass
at which the transition occurs. 
It is dependent, however, on the strength of the transition, $\eta$, as well
as on the exact time at which the fluctuation crosses into the horizon, in
particular, if shortly before onset, during, or shortly after completion of the
transition. This \lq\lq time\rq\rq\ may be parameterized by 
the ratio of cosmic average energy density at fluctuation horizon crossing
and some typical energy density at the transition,
$\tau_{hc} \equiv \epsilon_0(t_0)/\epsilon_h(T_c)$.
The threshold for PBH production is thus a function of 
$\tau_{hc}$ and $\eta$ only

\begin{equation}
\label{length1}
\delta_{\rm c}^{\rm FPT} =
\biggl({\delta\epsilon\over\epsilon}\biggr)_{\rm c,hc}\bigl(
\tau_{\rm hc}, \eta\bigr) \,\, .
\end{equation} 
In contrast, the resulting PBH masses for collapsing fluctuations are  
approximately determined by the horizon mass, 
$M_{\rm h}\approx (4\pi /3)\epsilon_{\rm h} R_{\rm h}^3
(\epsilon_{\rm h},\eta)$, at which the
transition occurs, and are only weakly dependent on $\eta$ and $\tau_{\rm hc}$. 
For a more accurate determination of the PBH mass
spectrum it is necessary to convolve the distribution function for 
the pre-existing density fluctuations, $P(\delta )d\delta$,
with a scaling relation for the resulting PBH masses,
$M_{\rm pbh} = kM_{\rm h}(\delta - \delta_{\rm c}^{\rm FPT})^{\gamma}$ as
shown in \cite{Nie2}. Here $k$ is a dimensionless constant.
This mass spectrum may be somewhat dependent on
$\eta$ and $\tau_{\rm hc}$.

In order to follow the hydrodynamic evolution of fluid elements at $T_{\rm c}$, 
on length scales much larger than
the mean separation between high- and low- energy density phase, it is
necessary to specify
an effective equation of state describing the mixture of phases in thermodynamic
equilibrium. Note that the assumption
of a near-to-equilibrium, first-order phase transition, 
implies negligible super -cooling and -heating and a typical mean
separation between phases much shorter than the cosmological horizon. These
assumptions are met if nucleation of new phase is efficient.
Since a fluctuation scale length is of the order of the cosmological horizon the
details of the distribution of phases on small scales should have
vanishing influence on the all-over fluctuation dynamics.  
The fluid is in phase mixture, for energy densities between 
$\epsilon_{\rm h}(T_{\rm c})$, where the volume fraction occupied by high-energy
density phase, $f_{\rm h}=1$, and, $\epsilon_{\rm l}(T_c)$, 
where $f_{\rm h}=0$. The
effective equation of state may thus be written 

\begin{equation}
\label{high}
p(\epsilon )={1\over 3}\bigl(\epsilon - L\bigr)\quad ;\quad v_{\rm s}^2={1\over 3}
\quad ;\quad \epsilon\geq \epsilon_{\rm h}(T_{\rm c})\,\ ;
\end{equation}

\begin{equation}
\label{mix}
p(\epsilon )={1\over 3}\epsilon_{\rm l}(T_{\rm c}) \quad ;\quad v_{\rm s}^2={0}
\quad ;\quad \epsilon_{\rm l}(T_{\rm c})  < \epsilon < \epsilon_{\rm h}(T_{\rm c})\,\ ;
\end{equation}

\begin{equation}
\label{low}
p(\epsilon )={1\over 3}\epsilon \quad ;\quad v_{\rm s}^2={1\over 3}
\quad ;\quad \epsilon\leq \epsilon_{\rm l}(T_{\rm c})\,\ ,
\end{equation}  
where $\epsilon_{\rm l}(T_{\rm c})$ and  $\epsilon_{\rm h}(T_{\rm c})$ 
are constants.
Note again that the pressure response, $v_{\rm s}$, of mixed phase is exactly zero
in thermodynamic equilibrium, whereas pressure itself is substantial.

\section{Results and Discussion}
\label{results}

We followed the evolution of density fluctuations upon horizon crossing
during a cosmological phase transition with scaled latent heat
$\eta =2$ (in particular, we chose
$\epsilon_{\rm h}(T_{\rm c}) = 1$ and $\epsilon_{\rm l}(T_{\rm c}) = 0.5$). 
Figures 1-3 show
results for a fluctuation with overdensity 
$(\delta\epsilon /\epsilon )_{\rm hc} = 0.535$ entering the cosmological horizon
at time $\tau_{\rm hc} = 1$, i.e. when the surrounding universe 
is at average energy density $\epsilon_{\rm h}(T_{\rm c})$ at the onset of
the phase transition. In Figure 1 we show the evolution of the radial energy
density profile of the fluctuation from the initial horizon crossing time $t_0$
to $20.1 t_0$. We choose constant proper time slicing, i.e. an
individual curve in Figure 1 represents the energy 
density of all fluid elements evaluated at identical proper time.
Energy density is shown as a function of scaled circumferential radius 
$R_{\rm sc}=(R/R_{\rm h}(t_0))(a_0/a)$ such that $R_{\rm sc}= constant$ 
for a fluid
element in simple FRW expansion. The horizon at $t_0$ is located
at $R_{\rm sc}=1$. The two horizontal dotted lines in Figure 1
indicate the regime of energy densities in which fluid elements exist within 
mixed phases.

\begin{figure}
\epsfxsize=80mm
\epsfbox{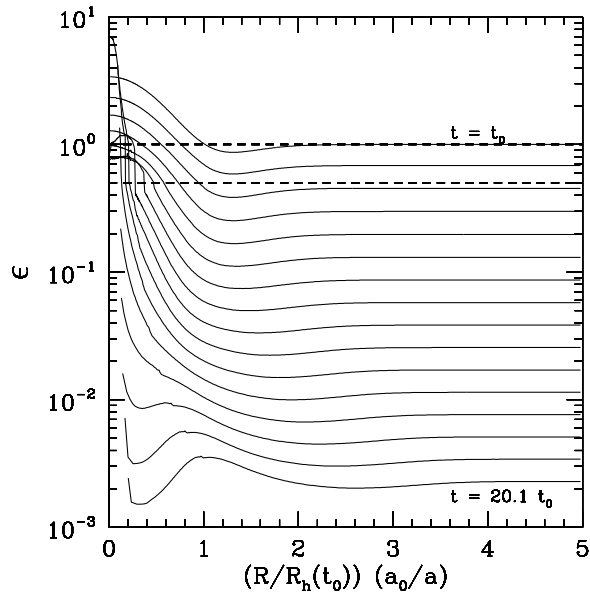}
\caption{\label{f1} Energy density, $\epsilon$, as a function of
scaled circumferential radius, $R_{\rm sc}=(R/R_{\rm h}(t_0))(a_0/a)$, for
a fluctuation with initial density contrast, $(\delta\epsilon /\epsilon)_{\rm hc}=
0.535$,
at horizon crossing. The initial horizon at $t_0$ is located at $R_{\rm sc}=1$.
From top to bottom, solid lines show the fluctuation at 1., 1.22, 1.49, 
1.82, 2.23, 2.72, 3.32, 4.06, 4.95, 6.05, 7.39, 9.03, 11.0, 13.5, 16.4, and 20.1 times the
initial time $t_0$. Constant proper time slicing was used. The horizontal
dashed lines indicate the energy densities at onset and completion
of the phase transition.
The formation of a PBH with $M_{\rm pbh}\approx 0.34M_{\rm h}(t_0)$ results.}
\end{figure}
\hspace{\fill}

The formation of a PBH with final mass $M_{\rm pbh}\approx 0.34M_{\rm h}(t_0)$
results from the evolution of the fluctuation shown in Figure 1-3.
Here we define the horizon mass at the initial time 
$M_{\rm h}(t_0) = \epsilon_0 V_{\rm h}(t_0)$. 
The fluctuation's selfgravity
exceeds pressure forces such that the fluctuation separates from Hubble flow
and
recollapses to high-energy densities at the center until an event
horizon forms.
Subsequent accretion of material on the young PBH continues until the
immense pressure gradients close to the event horizon launch an outgoing
pressure wave which significantly dilutes the PBH environment. Accretion
thereafter is negligible. The existence of a phase
transition facilitates the PBH formation process as is evident from Figure 2.
Figure 2 is a zoom into 
the core of the fluctuation shown in Figure 1.
For comparison, this figure also shows the evolution of a fluctuation with
the same initial conditions, but entering the cosmological horizon
during an ordinary 
radiation dominated epoch, by the dotted line. The strong pressure
gradients experienced by the fluctuation 
entering the horizon during an epoch
with equation of state $p=\epsilon /3$ prevent the formation of a PBH. 

\begin{figure}
\begin{minipage}{80mm}
\epsfxsize=80mm
\epsfbox{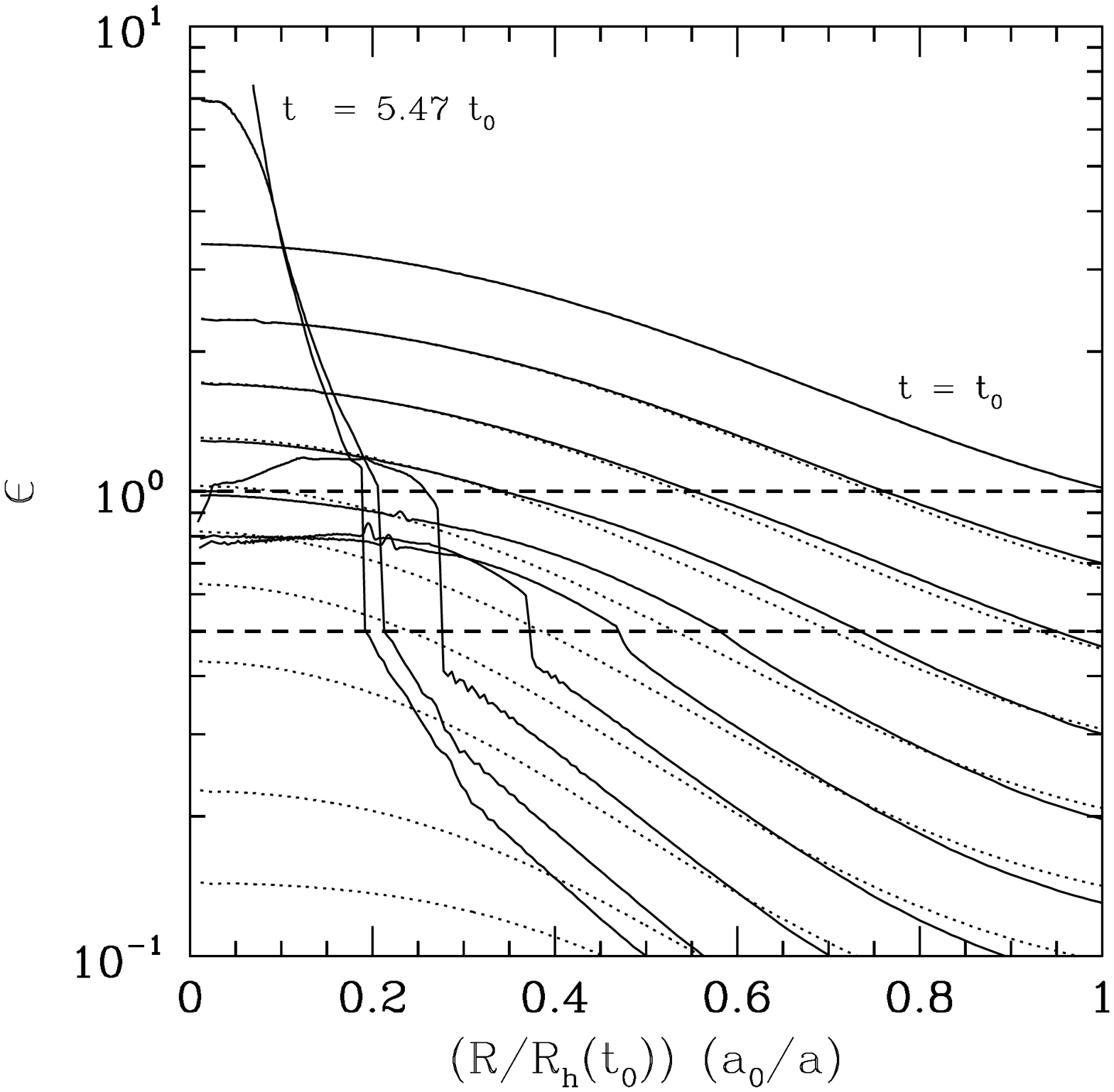}
\caption{\label{f2} A zoom into the central region of Figure 1. From top to
bottom, solid lines show the fluctuation at 1.0, 1.22, 1.49,
1.82, 2.22, 2.72, 3.32, 4.06, 4.95, and 5.47 times the
initial time $t_0$. The horizontal
dashed lines indicate the energy densities at onset and completion
of the phase transition. The dotted lines shows, for comparison, the evolution of
a fluctuation with the same initial fluctuation parameters, but entering the
cosmological horizon during an epoch with equation of state $p =\epsilon /3$.}
\end{minipage}
\hspace{\fill}
\begin{minipage}{80mm}
\epsfxsize=80mm
\epsfbox{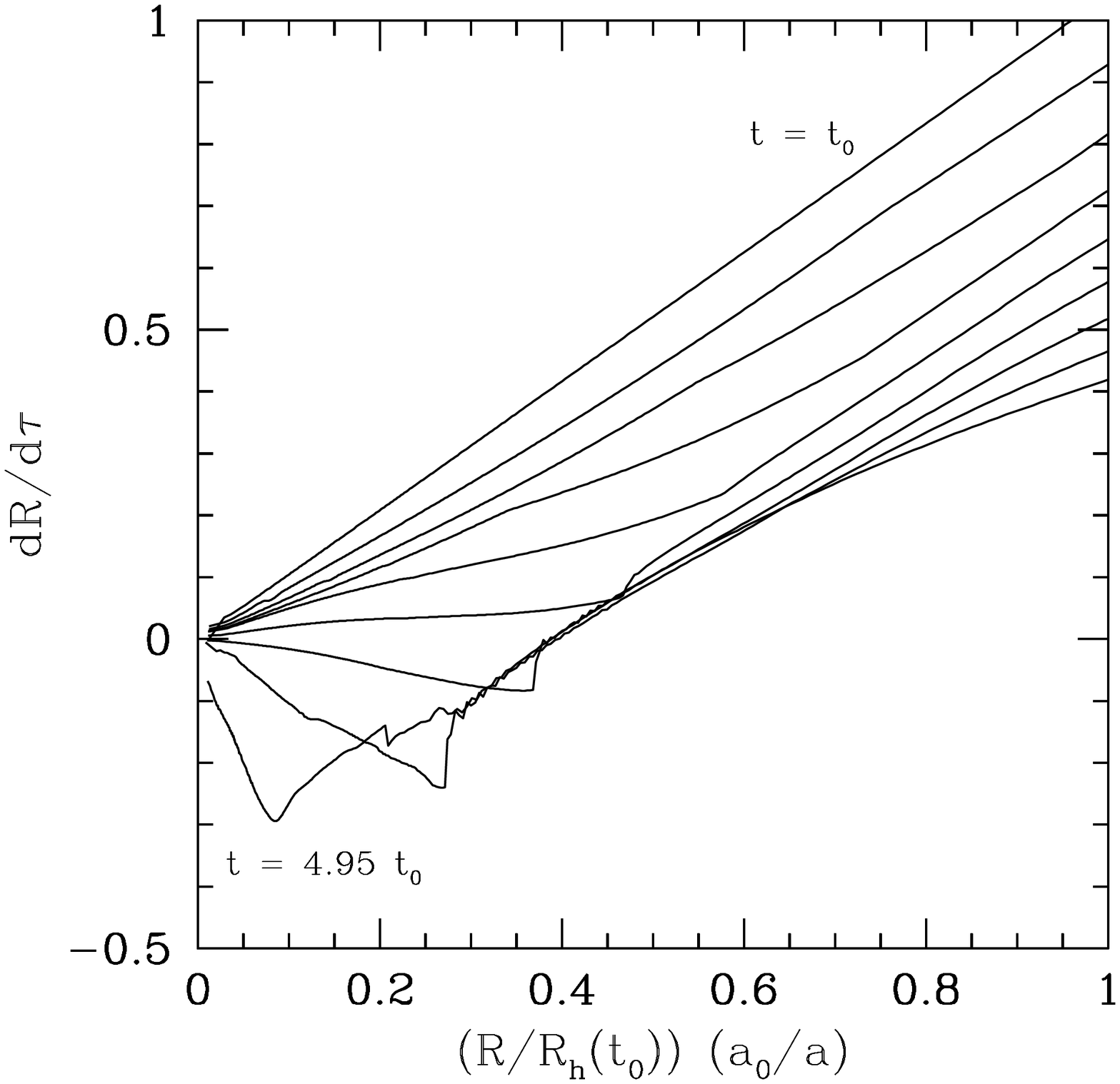}
\caption{\label{f3} Coordinate velocity, $\partial R/\partial\tau$, with $\tau$
proper time of fluid elements, as a function of scaled circumferential
radius, $R_{\rm sc}=(R/R_{\rm h}(t_0))(a_0/a)$. 
The solid lines show, from top to bottom,
the coordinate velocities at times 1.0, 1.22, 1.49,
1.82, 2.22, 2.72, 3.32, 4.06, and 4.95 times $t_0$ for the same fluctuation
and equation of state as shown in Figures 1 and 2 by the solid lines.}
\end{minipage}
\end{figure}

The evolution of the fluctuation in the presence of a phase transition
proceeds as follows. The initial phase 
between times $t_0$ and $\sim 2t_0$ is characterized by a core in high-energy
density phase surrounded by an envelope in mixed phase. Cosmological expansion
results in the decrease of energy densities in core and envelope. 
Nevertheless, the
fluctuation's overdensity decelerates the material with respect to FRW
expansion. This is evident from Figure 3, which
shows the coordinate velocity $\partial R/\partial\tau$
for the fluctuation shown in Figures 1 and 2, with $\tau$ proper
time. In unperturbed Hubble flow, coordinate velocities would be straight lines
with decreasing slope for increasing cosmic time. The deceleration of the
expanding fluid is stronger for those fluid elements existing in mixed phase
due to the absence of pressure gradients. Material in the core remains in
high-energy density phase until $t \approx 2.2$. 
Its deceleration is first stronger than FRW due to selfgravity until at
$t\approx 2t_0$, the increasing pressure gradient begins to counteract
the collapse. 
The pressure gradient would give rise to the launching of a pressure
wave and the dispersion of the fluctuation if core and envelope wouldn't
join at $t\approx 2.2t_0$ and exist in mixed phase. 
In the subsequent phase between
times $\sim 2.2t_0$ and $\sim 4t_0$, a distinctively separated core
develops. It is 
distinct through discontinuities in energy density and coordinate velocity.
The core first undergoes homologous, but sub-Hubble, expansion which
turns into homologous contraction at $t\approx 3.3t_0$. 
The core energy density is almost
uniform and increases with time. Initially the core mass-energy
still reduces somewhat. Shells on the boundary of the
core are expelled when they experience strong
pressure gradients as their energy density falls below $\epsilon_{\rm l}$.
In the final phase of collapse of the core region between $\approx
4 t_0$ and $\approx 5t_0$
the fluid exists exclusively in high-energy density phase surrounded by
fluid in low-energy density phase. 
The collapse proceeds no further
homologously, rather, inner parts of the core contract faster than outer parts.
At $t\approx 5t_0$ 
an event horizon forms.
The resulting young PBH rapidly increases its mass 
up to $M_{\rm pbh}\approx 0.06M_{\rm h}(t_0)$ 
at $t\approx 5.5t_0$. 
Subsequent slow long-term
accretion onto the PBH increases the PBH mass further to 
$M_{\rm pbh}\approx 0.34M_{\rm h}(t_0)$.  

The dynamics of PBH formation during a first-order phase transition
is different from the one experienced during ordinary radiation dominated
epochs. In the latter there is no obvious segregation between collapsing
fluctuation and expanding universe, in particular, there are no discontinuities.
Further, the collapse of fluctuations during ordinary radiation dominated
epochs does not proceed in a homologous fashion. Figures 2 and 3 reveal the
existence of instabilities in the evolutionary calculations of the PBH formation
process. It may be observed that material within mixed phase develops a 
small-scale perturbation of growing amplitude. This is not surprising, since
with $v_{\rm s} = 0$ in mixed phase the fluid is
Jeans-unstable to collapse on all scales. The perturbation extends
over many zones and its amplitude and size are 
unchanged with an increase of numerical
viscosity and/or number of zones. We note that there may be other
physical instabilities 
for $v_{\rm s} = 0$ which we, however, do not
observe due to our choice of initial conditions. It has been argued that an
instability of non-gravitational origin exists for sound waves which experience
a sudden drop in the speed of sound \cite{Schmid1,Schmid2}.  
Finally, we also observe a numerical instability of the adopted algorithm
which occurs when the energy density of fluid shells falls slightly below
$\epsilon_{\rm l}(T_{\rm c})$. 
Such shells experience immense acceleration when leaving
the regime of mixed phase with large energy density gradients. The amplitude
of the perturbations may be regulated by our choice of numerical viscosity,
nevertheless, the final black hole mass is virtually unaffected
by an increase/decrease of numerical viscosity.

\begin{figure}
\epsfxsize=80mm
\epsfbox{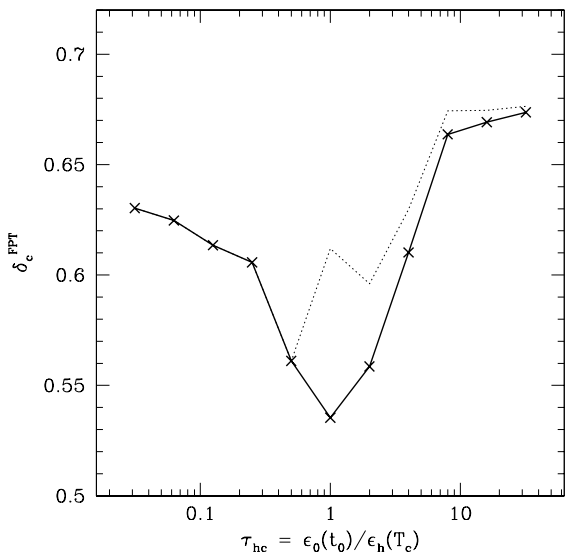}
\caption{\label{f4} Energy overdensity threshold for PBH formation,
$\delta_{\rm c}^{\rm FPT}$ (solid line), 
for fluctuations entering the cosmological horizon during, or close
to a first-order phase transition, as a function of horizon crossing time,
$\tau_{\rm hc}=\epsilon_0(t_0) /\epsilon_{\rm h}(T_{\rm c})$. 
The energy densities at the
onset and completion of the transition are chosen $\epsilon_{\rm h}(T_{\rm c}) =1$
and $\epsilon_{\rm l}(T_{\rm c}) =0.5$, respectively. 
The crosses are points determined
from numerical simulation. The solid line is an interpolation between crosses.
The dotted line shows $\delta_{\rm c}^{\rm FPT}/(1+w)$,
with $w$ the cosmic average
$p/\epsilon$ at horizon crossing of the fluctuation. (See text for
further explanations).} 
\end{figure}
\hspace{\fill}

The main result of our study is displayed in Figure 4. This figure shows
the threshold for PBH formation, $\delta_{\rm c}^{\rm FPT}$, for a phase transition
with scaled latent heat, $\eta =2$, as a function of the horizon crossing time
of the fluctuation, $\tau_{\rm hc}$, by the solid lines. Crosses represent the
lowest $(\delta\epsilon /\epsilon)_{\rm hc}$ for which PBH formation results.
The relative accuracy in $\delta_{\rm c}^{\rm FPT}$ 
is estimated to be on the 1\% level.
It is evident that over a range of horizon crossing times the energy overdensity
required for PBH formation is below that for PBH formation during ordinary
radiation dominated epochs. It should be clear from the introductory 
remarks that the slightest bias in favor of forming PBH during a phase
transition may imply that essentially all PBH are formed on the scale
associated with minimum $\delta_{\rm c}^{\rm FPT}$.
However, assuming a Harrison-Zel'dovich, 
exactly scale-invariant spectrum for the underlying density perturbations,
the amplitudes of energy density perturbations in uniform
Hubble constant gauge upon horizon crossing are not {\it exactly}
constant during
epochs with varying equation of state. Rather, modes with varying length scale
cross into the horizon with equal
$\xi = (1+w)^{-1}(\delta\epsilon /\epsilon)_{\rm hc}$
\cite{KT90}. Here $w = p/\epsilon$,
such that for $w$ smaller than $1/3$, as in Eq. (\ref{high}) - Eq. (\ref{low})  
for $\epsilon > 
\epsilon_{\rm l}(T_{\rm c})$, the horizon crossing amplitude  
$(\delta\epsilon /\epsilon)_{\rm hc}$ of perturbations is reduced.
This partially decreases the bias to form PBH during phase transitions.  
In Figure 4 the dotted line shows a $\tilde{\delta}_{\rm c}^{\rm FPT}$ 
which corrects
for this effect. Remarkably, the combined effects of reducing the threshold
for PBH formation by the dynamics of fluctuations during a first-order
transition and the decrease of the horizon crossing amplitude of energy density
fluctuations due to the changed equation of state, result in a double
dip structure. 

We have also computed test runs for fluctuations entering the horizon during
a phase transition with $\eta =10$. Our results indicate that with increased
strength of the phase transition $\delta_{\rm c}^{\rm FPT}$ is further
decreased. 
With the discovery of critical phenomena in general relativity it has been
recognized that the resulting black hole masses in nearly
critically collapsing space-times
obey scaling relations \cite{crit}, such as 
$M_{\rm pbh} = K(\delta - \delta_{\rm c})^{\gamma}$. For a given
matter model, $\gamma$ is independent of initial conditions, while $K$
depends on the specific choice of the control parameter
$\delta$. Specifically, a collapsing radiation fluid has
$\gamma \approx 0.36$ \cite{gamma}. In our case,
$\delta$ may be associated with $(\delta\epsilon /\epsilon)_{\rm hc}$
and $K$ with $kM_{\rm h}$. We have explicitly verified that 
$\gamma \approx 0.36$ also holds for the cosmological PBH formation
process during 
ordinary radiation dominated epochs \cite{Nie1}. We have attempted to derive
a scaling exponent $\gamma$ for the PBH formation process during
epochs with equation of state given by Eq. (\ref{high}) - Eq.
(\ref{low}). Our numerical simulations do not clearly indicate a
simple scaling relation as above. 
Preferred values of $\gamma$ were rather large
(i.e. $\gamma > 2$), even though somewhat dependent on the range
of black hole masses to which we fit the scaling relation. 
In contrast to our work in \cite{Nie1}, convincing 
verification of a scaling relation and a
value for the exponent $\gamma$ would require the use of adaptive mesh
techniques. This would allow to simulate the formation of PBH with masses far
smaller than $0.1M_{\rm h}$. We note that convolving
a Gaussian density perturbation probability function with the
preferred scaling relation of our numerical simulations would predict
average masses $M_{\rm pbh}$ far below $0.1 M_{\rm h}$. Clearly, resolution
of this issue would be desirable. 

\section{Conclusions}

We have performed hydrodynamic general-relativistic
simulations of the PBH formation process when initially super-horizon
energy density fluctuations enter the cosmological horizon during a first-order
phase transition. We have verified that the significantly diminished pressure
forces for fluid undergoing the phase transition facilitate the PBH formation
process. We have shown that the threshold of the 
amplitude of energy density fluctuations above which PBH formation results
is smaller for fluctuations crossing into the horizon during a first-order
phase transition than for fluctuations 
crossing into the horizon when the
equation of state is ordinary (i.e. $p =\epsilon /3$). PBH formation 
from pre-existing adiabatic fluctuations is
more probable during first-order transitions, even when 
the reduction of fluctuation amplitudes of Harrison-Zel'dovich type
which cross into the horizon for fluid pressures smaller than $\epsilon /3$, 
is taken
into account. Our simulations could not clarify if simple scaling relationships 
with scaling exponent $\gamma \approx 0.36$ for
the resulting PBH masses, as found in many studies of critical phenomena in
general relativity, also apply to the PBH formation process during cosmological
first-order phase transitions. Our simulations favor much larger scaling
exponents, possibly yielding small typical PBH masses (in units of the
horizon mass). Further study with adaptive mesh techniques is required to
resolve this issue. 

Our results may have cosmological implications for the
PBH mass spectrum. It was so far believed that PBH formation 
from pre-existing adiabatic perturbations in the early
universe is equally likely on all scales,
only dependent on the statistics of the pre-existing density perturbations. 
The dynamics of fluctuations during cosmological phase transitions may easily
introduce a strong enough bias to form PBH only on the approximate 
horizon mass scales during epochs with phase transitions. This may
yield a highly peaked mass function for PBHs.
Here the term cosmological phase transition
may be loosely interpreted, since a reduction of pressure forces is not only
anticipated during first-order phase transitions, but may also occur
during higher-order transitions, cross-overs, and particle annihilation periods.
A prime candidate for such a transition is QCD color-confinement due to
the large number of degrees of freedom participating in the transition.
Whether, or not, PBHs have been produced in the early universe in
cosmologically interesting numbers remains an open question.

We wish to thank T. Baumgarte for providing the original version of
the hydrodynamical code, and T. Abel, E. M\"uller, and A. Olinto
for helpful discussions. JCN acknowledges the support of an
Enrico-Fermi-Fellowship.

\end{document}